\documentclass[12pt,reqno]{article}
\usepackage{amsmath,amssymb,amsthm}
\usepackage{comment}

\usepackage[T2A,T1]{fontenc}
\usepackage[utf8]{inputenc}
\usepackage[russian,english]{babel}

\usepackage[text={6in,10in},centering]{geometry}

\usepackage[shortlabels]{enumitem}
  \setlist[enumerate,1]{leftmargin=25pt}
  \setlist[itemize,1]{leftmargin=25pt}
  \setlist[description,1]{leftmargin=15pt}

\usepackage{graphicx}
\usepackage[font=small]{caption}
\usepackage{floatrow}
\usepackage{subcaption}

\usepackage{hyperref}
\usepackage{mathabx}
\usepackage{quoting}
\usepackage{url}

\newenvironment{Blue}{\noindent\color{blue}}{}

\newenvironment{q}
 {\begin{quoting}[vskip=6pt,leftmargin=\parindent,rightmargin=0pt]}
 {\end{quoting}}

\newenvironment{Red}{\noindent\color{red}}{}

\theoremstyle{definition}

\renewcommand\a{\ensuremath{\forall}}
\newcommand\Ar{\smallskip\noindent\textbf{A: }}
\newcommand\bs{\ensuremath{\,\big/\,}}

\renewcommand\phi{\ensuremath{\varphi}}
\newcommand\qef{\hfill$\triangleleft$} %Quod erat faciendum
\newcommand\Qn{\smallskip\noindent\textbf{Q: }}
\newcommand\set[1]{\ensuremath{\left\{#1\right\}}}

\title{The umbilical cord of finite model theory}
\author{Yuri Gurevich\\
\small Computer Science \& Engineering, University of Michigan}
\begin{document}
\date{}
\maketitle
\thispagestyle{empty}

\begin{abstract}
Model theory was born and developed as a part of mathematical logic.
It has various application domains but is not beholden to any of them.
A priori, the research area known as finite model theory would be just a part of model theory but didn't turn out that way.
There is one application domain --- relational database management --- that finite model theory had been beholden to during a substantial early period when databases provided the motivation and were the main application target for finite model theory.

Arguably, finite model theory was motivated even more by complexity theory.
But the subject of this paper is how relational database theory influenced finite model theory.

This is NOT a scholarly history of the subject with proper credits to all participants.
My original intent was to cover just the developments that I witnessed or participated in.
The need to make the story coherent forced me to cover some additional developments.
\end{abstract}

\section{Prelude}\label{s:intro}

\textbf{Q}\footnote{Quisani is my former student.}:
How come finite model theory is computer science while model theory is mathematics?

\noindent\textbf{A}\footnote{The author}:
Let me think.
Model theory is a part of mathematical logic which was born and developed as a part of foundations of mathematics.
Finite model theory (FMT) was developed much later.
It did not exist as a separate research area before the 1970s.
FMT was developed primarily by computer scientists.
It was --- and is --- much influenced by complexity theory which is viewed traditionally as theoretical computer science (though mathematicians often view it as mathematics).
Also, during a substantial early period FMT was much influenced by (relational) database theory (DBT), bona fide computer science.
I doubt that FMT would have become a recognizable research area without that  DBT link.
In that sense, the DBT link is the umbilical cord of FMT.

\Qn Tell me about that influence of DBT on FMT.

\Ar I'll try my best.
But let me make two reservations.
I am not a DBT expert and, more importantly, I am a wrong person to render a scholarly history of the subject complete with proper credits to all participants.
It takes a special talent to give a scholarly account of a research field; I don't have such talent.
I will tell you primarily about the developments that I witnessed or participated in.

\Qn One last question before you start your story.
I presume that logicians studied finite models before the 1970s?

\Ar Yes, they did. Let me start with that logic prehistory of FMT.

\section{Prehistory}\label{s:pre}

Finite structures were relevant to humanity long before the notion of structure was introduced, in fact before the notion of theorem was introduced.
Consider for example the fact that a finite set can be counted in any order, always giving the same result.
The special cases of this fact --- for 2-element sets, 3-element sets, etc. --- must have been known very early.
The Greeks probably knew the general fact before the notion of theorem was introduced.

While model theory was studied (e.g.\ by Skolem and L\"owenheim) in the beginning of the 20th century, finite model theory arguably became a separate research area in the 1970s, and the term ``finite model theory'' was coined only toward the end of the century.

As the rest of this section shows, finite model theory was studied and used before it became a separate research area.

\subsection{Boris A. Trakhtenbrot}\label{sb:trakh}

In the paper \cite{Trakh1950} Boris Trakhtenbrot proved that finite satisfiability is not decidable. There is no algorithm that, given a first-order sentence $\phi$, decides whether $\phi$ has a finite model.

It follows that finite validity is not recursively enumerable, that is the set of finitely valid (valid on all finite models) first-order sentences is not recursively enumerable.
Indeed, let $S$ be the set of finitely valid sentences and $S'$ the complement of $S$.
Sentences in $S'$ have finite counterexamples.
It follows that $S'$ is recursively enumerable.
If $S$ were recursively enumerable as well, then the following (not very feasible) algorithm would decide, given a sentence \phi, whether it is in $S$ or in $S'$.
Keep enumerating a list of $S$ sentences and a list of $S'$ sentences until \phi\ occurs in one of the lists.

The failure of recursive enumerability is important. It means that there is no reasonable deductive system for finite validity.
This contrasts with the situation for general validity (validity in arbitrary, not necessarily finite, structures) where there is such a deductive system.

In the paper \cite{Trakh1953} published in 1953, Trakhtenbrot strengthened his result. The set of finitely satisfiable sentences and the subset of unsatisfiable sentences are recursively inseparable: There is no recursive set that contains one of the two sets and is disjoint from the other.

\begin{figure}[H]
\includegraphics[width=10cm]{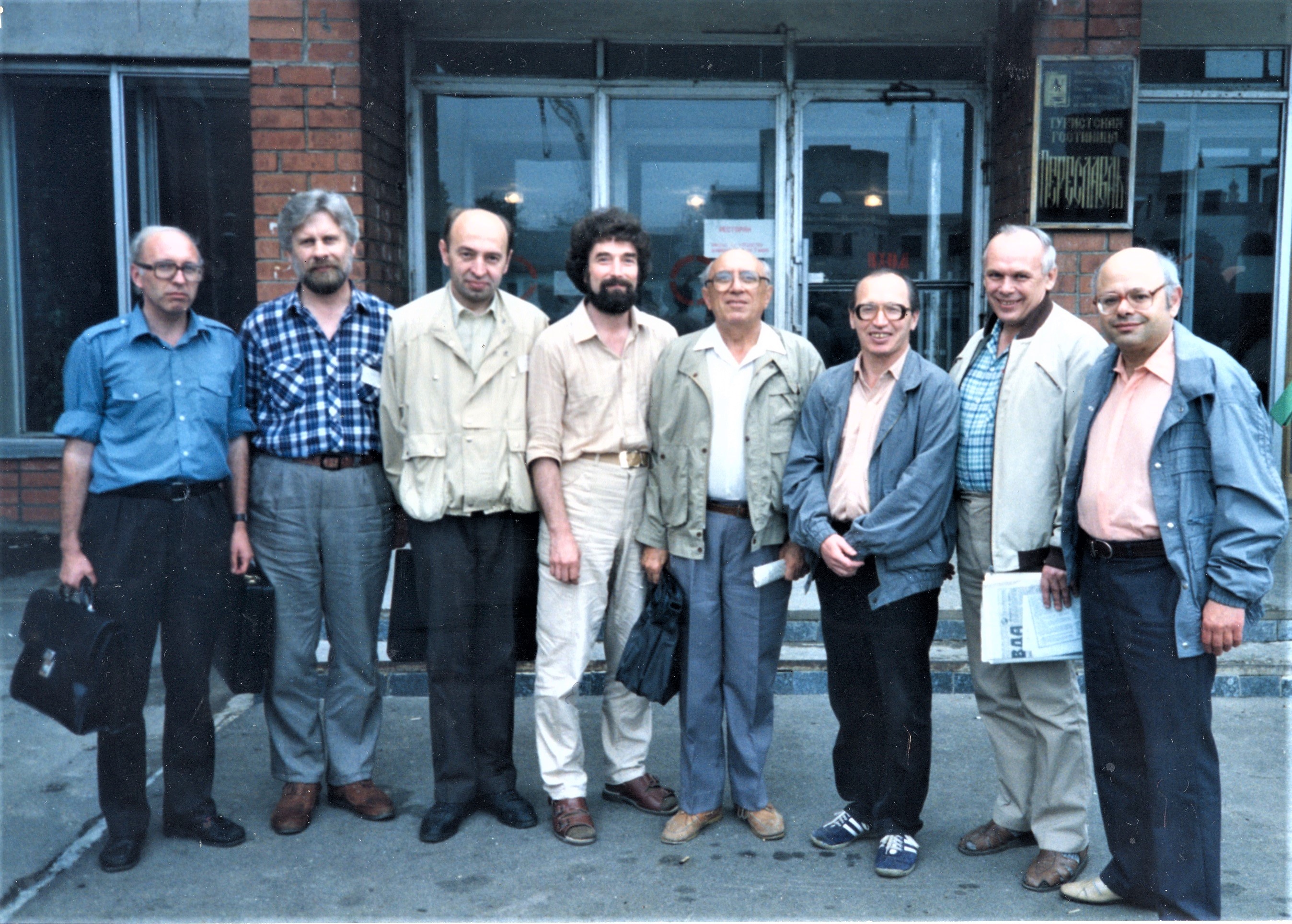}
\caption{\footnotesize
Boris Trakhtenbrot with some of his former students in 1992.
Left to right: Vladimir Sazonov, Janis Barzdins, Rusins Freivalds, Alexandre Dikovsky, Boris Trakhtenbrot, Mars Valiev, Miroslav Kratko, and Valery Nepomnyaschy. %(Courtesy of the Trakhtenbrot family archive.)
}
\label{Trakh}
\end{figure}

Since FMT is typically considered as a part of theoretical computer science, it may be fitting to note that Trakhtenbrot was also a pioneer of computer science and had many good students (Figure~\ref{Trakh} shows him with a few of them). See also \cite{ADR} in this connection.

I first met Trakhtenbrot in the mid-1960s in the Novosibirsk Academgorodok \cite{W-Akademgorodok} where he was running a lively seminar on the nascent computer science. In spite of the difference in age, we became friends. I used to come to the Akademgorodok from time to time to give a talk at the Algebra \& Logic seminar run by academician Anatoly Maltsev. On the same occasions, I would attend Trakhtenbrot's seminar and talk to Trakhentbrot. I remember how he introduced time and space complexity (by means of so-called ``signalizing functions'') independently of his Western colleagues. I could have become a computer scientist much earlier than I did. But, at the time, I was moving from algebra to logic and it was logic that fascinated me.

\subsection{Dana Scott, Patrick Suppes, and William W. Tait}
\label{sb:SST}

\begin{figure}[H]
  \begin{subfigure}[b]{0.3\textwidth}
    \includegraphics[width=.75\textwidth]{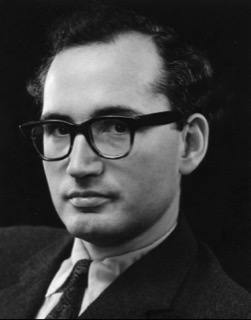}
    \caption{Dana S. Scott\mbox{\hspace{25pt}}}
    \label{Scott}
  \end{subfigure}
  \begin{subfigure}[b]{0.3\textwidth}
    \includegraphics[width=.85\textwidth]{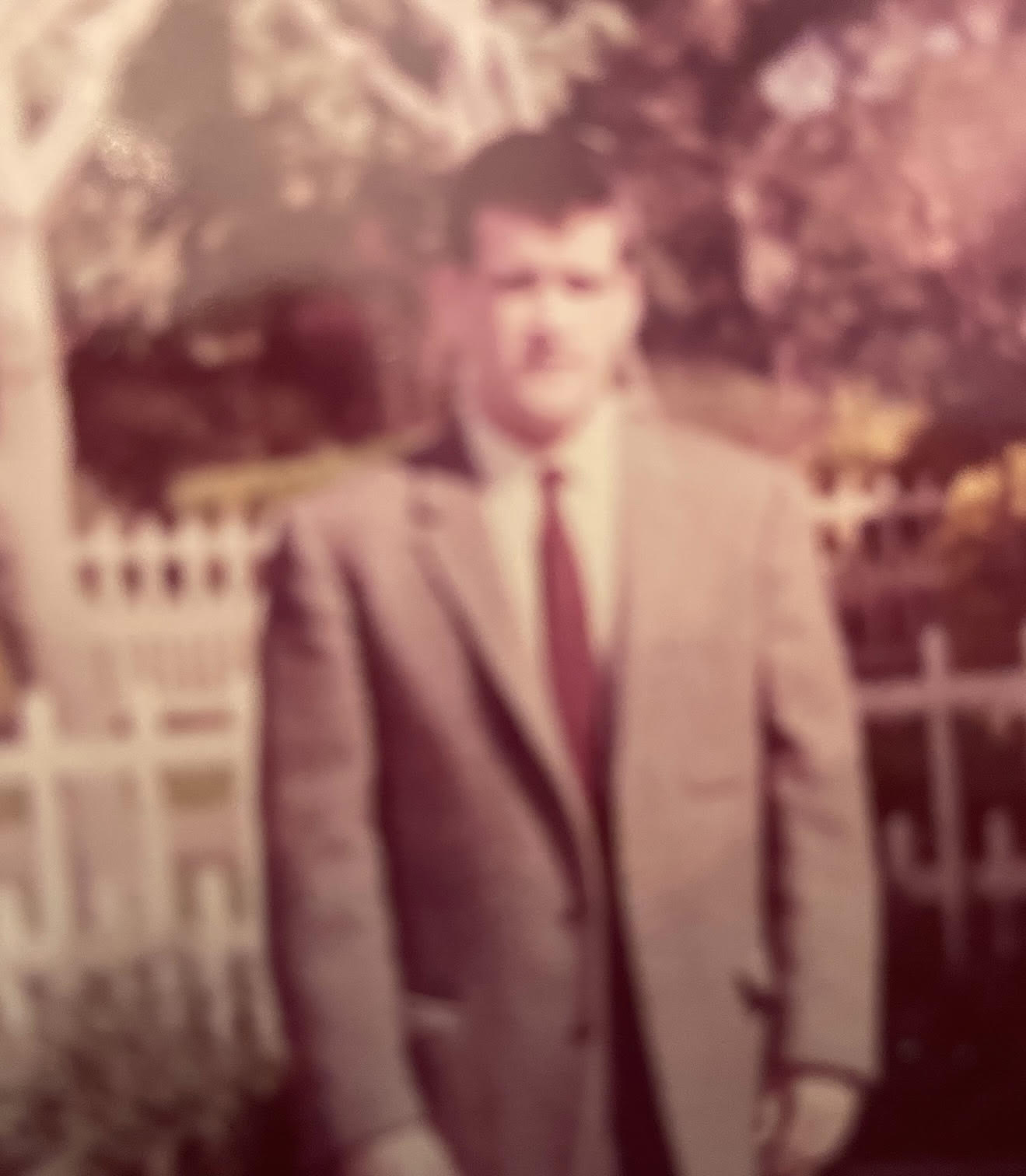}
    \caption{William W. Tait\mbox{\hspace{16pt}}}
    \label{Suppes}
  \end{subfigure}
 \begin{subfigure}[b]{0.3\textwidth}
   \includegraphics[width=.96\textwidth]{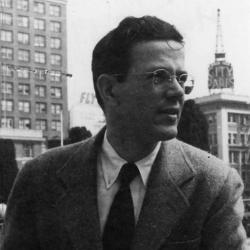}   \caption{Patrick C. Suppes\mbox{\hspace{8pt}}}
   \label{Suppes}
 \end{subfigure}
\end{figure}

It is easy to see that a purely universal sentence, that is a sentence of the form
\[ \phi = \a x_1 \a x_2 \dots \a x_n \Phi(x_1,x_2,\dots,x_n) \]
where $\Phi$ is quantifier-free, is preserved by submodels: If $A\models \phi$ and $B$ is a submodel of $A$ then $B\models \phi$.
According to the \L o\'s-Tarski theorem \cite[Theorem~5.2.4]{CK}, there is a converse: Any sentence that is preserved by substructures is equivalent to a universal sentence.

In the 1958 paper \cite{Suppes}, Dana Scott and Patrick Suppes conjectured that the \L o\'s-Tarski theorem remains valid when only finite structures are taken into account.
In the 1959 paper \cite{Tait}, William W. Tait disproved their conjecture.
In retrospect, his counterexample is rather simple.
Essentially it is a first-order sentence expressing that a linear order with a first element has also a last element.
Every finite model has the property, and so the property is preserved by submodels of finite models.
But an infinite model may violate the property in spite of being a submodel of one that has the property;
the smallest example is the linear order $\omega$ of natural numbers as a submodel of $\omega+1$.

But there was more in the Scott-Suppes article than the conjecture mentioned above.
We will return to their paper later in this section.

\subsection{Yuri Glebsky and his students}\label{sb:01}

In the 1966 International Congress of Mathematicians in Moscow, Yuri Glebsky announced an unexpected zero-one law for first-order logic \cite{Glebsky}.

\noindent
\begin{minipage}{.45\textwidth}
Consider a finite set $V$ of relation symbols and restrict attention to models of vocabulary $V$ with the universe of the form \set{1, 2, \dots, n}.
For each first-order sentence \phi\ in vocabulary $V$, let $B(n)$ be the total number of $n$-element models and $A(n)$ the number of such models satisfying $\phi$.
The zero-one law states that, for every $V$ and \phi\ as above, $\lim_{n\to\infty} A(n) \bs B(n)$ exists and is equal to either 0 or 1.
\end{minipage}
\begin{minipage}{.1\textwidth}
\mbox{}
\end{minipage}
\begin{minipage}{.4\textwidth}
\begin{figure}[H]
\includegraphics[width=125pt]{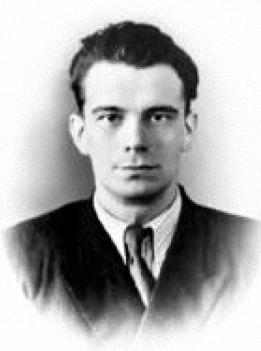}
\caption{Yuri V. Glebsky}
\label{Glebsky}
\end{figure}
\end{minipage}

In 1969, Glebsky and three of his students --- Dmitry I. Kogan, Mark I. Liogonky, and Vladimir A. Talanov --- published a proof of the zero-one law \cite{Glebsky1969}.
In 1972, the paper was duly translated to English but wasn't noticed in the West.
By chance, I knew paper \cite{Glebsky1969}.
This wasn't because I was so scholarly; that has never been my forte.
But, in 1970, I served as an official opponent at the public defense of the PhD thesis of Liogonky.
In the thesis, Liogonky proved that the 0-1 law remains valid if the structures are counted up to isomorphism

In 1972, Ron Fagin announced the 0-1 law, which he had discovered independently in the meantime \cite{Fagin1972}.
He gave an elegant proof of the law in his thesis \cite{Fagin1973} and published it in \cite{Fagin1976}.
A number of years later, I met Fagin and told him about Glebsky.
Fagin was surprised, but he found the English translation of the 1969 paper, and the matter was settled.

In 1977 Glebsky drowned in the Volga trying to save his son.
His students admired him, and his untimely death devastated them  \cite{LT}.

\subsection{Other relevant results}

\Qn To what extent is the collection of relevant ``prehistoric'' results above exhaustive?

\Ar It is definitely not exhaustive.
There are many pre-1970 results that are relevant to some degree.
For example, the recursive inseparability result of Trakhtenbrot provoked more subtle recursive inseparability results in algebra and logic.
In the book \cite{G000}, an appropriate reduction theory is presented, and it is proven that, for many natural classes $K$ of first-order sentences, the set of finitely satisfiable $K$ sentences and the set of unsatisfiable $K$ sentences are recursively inseparable.

Also there is at least one very relevant result that is missing above.
Commenting on a draft of this paper, Erich Gr\"adel reminded me of the B\"uchi-Elgot-Trakhtenbrot Theorem \cite{Buchi, Elgot, Trakh1962}, which he views as the first  descriptive complexity result.
The theorem states that a set of strings in any finite alphabet is accepted by a finite-state automaton if and only if it is definable in monadic second-order logic (MSO).
It is certainly the first descriptive complexity result that I know.

\section{Database theory}\label{s:codd}

Edgar Frank ``Ted'' Codd wouldn't get into a model theory hall of fame.
His famous theorem, known simply as Codd's Theorem \cite{W-Codd's_theorem}, is good model theory, but it wasn't real news to logicians as we explain below.
Yet he is a hero of our story.
He invented the relational model for database management, which facilitated the birth of finite model theory as a separate research area.

\begin{figure}[H]
\includegraphics[width=110pt]{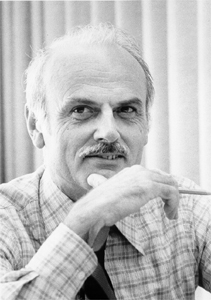}
\caption{Edgar F. Codd}
\label{Codd}
\end{figure}
It may be hard for a theorist with no practical experience to appreciate Codd's contribution.
Codd observed that databases can be adequately represented by (first-order) structures \cite{Codd1970,Codd1972}.
But structures had originally been designed to represent data, or knowledge.

\begin{q}
\Qn Designed by whom?

\Ar By architects of what is now called model theory, starting from L\"owenheim and Skolem, including G\"odel and Tarski, especially Tarski; see \cite{Tarski1935} for example.

\Qn Mathematical logic was developed within the framework of foundations of mathematics.
Only mathematical data might have been of interest to logicians.

\Ar I don't know what you mean by mathematical data.
Logicians certainly did not restrict attention to numerical data.
Any data whatsoever may be involved in a mathematical problem and thus become ``mathematical data'' of interest to logicians.

Besides, logicians have been interested in problems beyond foundations of mathematics.
The Scott-Suppes paper mentioned in \S\ref{sb:SST} is on the foundations of theories of measurements. The authors foresaw relational databases:
\begin{q}
``From an abstract standpoint, a set of empirical data consists of a collection of relations between specified objects \dots\
we treat sets of empirical data as being (finitary) relational systems'' \cite[\S1]{Suppes}. \qef
\end{q}
\end{q}

In article \cite{Codd1972}, Codd defined a variable-free \emph{relational algebra} and a \emph{relational calculus}.
The latter is a form of first-order logic, with propositional connectives and quantifiers, which is convenient for his purposes.
He also proved a theorem, Codd's Theorem, establishing that relational algebra has essentially the expressive power of first-order logic.

\begin{q}
\Qn Why essentially and not literally?

\Ar Database queries should be safe in the appropriate technical sense
\cite[\S5.3]{AHV}.
For example, no query should express the complement of a finite subset of an infinite domain.
Codd's Theorem establishes that relational algebra has the expressive power of the safe fragment of first-order logic.
\end{q}

But Codd's Theorem wasn't real news for logicians either.
An algebraic form of first-order logic, using so-called cylindric algebras, had been developed earlier by Tarski and his collaborators \cite{Tarski1948,Tarski1952,IL}.

\begin{q}
\Qn If logicians knew all that, then what made Codd's work important?

\Ar In one word, engineering.
There is a huge difference between treating sets of empirical data as relational systems and practical database management.
Codd's approach was practical, and the result was a model of database management superior to all competitors.
He received the Turing Award in 1981.

\Qn What are the benefits of the relational database management?

\Ar One is simplicity; all information is represented by data values in addition to a few relation symbols.
Another important benefit is clear and unambiguous semantics.
Yet another important benefit is that the relational approach admits high level query languages.

\Qn If we stick to theory, rather than practice, was there anything new in Codd's approach?

\Ar Yes, there was.
Normally, in logic and in mathematics in general, structures are static. You study a particular structure, e.g.\ the field of real numbers, or a class of structures, e.g.\ commutative groups.
But databases evolve and in that sense they are dynamic.
For example, consider the salary database for some organization.
New employees are hired, some employees leave the organization, some get raises, etc.
To incorporate evolution, the language of databases should allow us to update databases in various ways.

\Qn Did  you know Codd?

\Ar No; I wish I did.
Codd wrote his PhD thesis at the University of Michigan, but that was in the 1960s, long before I arrived there.
But I heard plenty about Ted Codd from his advisor John Holland.
John told me that Codd compromised his health trying to promote the relational approach at IBM.
His fight at IBM is described in many places, even in the Wikipedia article \cite{W-Codd}.
\end{q}

\section{The beginning}

\Qn After all these preliminaries are you ready to discuss FMT per se?
If yes, start with the beginning.

\Ar I am ready to discuss FMT but starting with the beginning is a challenge.
Until 1982, when I joined the University of Michigan and officially became a computer scientist, I was a mathematician and very different things were on my mind, primarily model theory.
So, preparing for this discussion, I exchanged letters with experts.
In particular, I posed two questions to Ron Fagin:

\begin{minipage}{\textwidth}
\begin{enumerate}[itemsep=0pt,leftmargin=0pt,topsep=10pt]
\item What do you count as the beginning of FMT? Is it your thesis?
\item Who coined the term ``finite model theory''? And when?
\end{enumerate}
\end{minipage}\\[10pt]
``Although Trakhtenbrot’s Theorem was the first important result,'' replied Ron on Nov.~9 2022, ``[I] do feel that my Ph.D. thesis launched FMT as a field.  Since I am biased, I called Moshe [Vardi] (whom I am cc-ing). He says he completely agrees that Trakhtenbrot had the first major result and that my thesis started FMT as a field.''

\noindent
A couple of days later, Moshe sketched for me his view of the early history.
\begin{q}
``In the 1970s, there are three independent lines of work:
\begin{enumerate}
\item[a.] 1970: Codd's paper --- start of relational DBT.
\item[b.] 1973: Fagin's dissertation --- `Contributions to the model theory of finite structures', where he proved $\Sigma_1^1 = \text{NP}$, the 0-1 Law, and that reachability is not definable in [existential] MSO.
\item[c.] 1980: Immerman's dissertation --- `First order expressibility as a new complexity measure.'
\end{enumerate}
The work in the early 1980s built on these three legs, even though it was not explicitly FMT.
For example, the 1982 papers by Immerman and by Vardi
built on these legs, but did not emphasize the finiteness feature.
Your papers in the 1980s raised the issue of finiteness and helped FMT
become a well-defined field.'' \hfill --- Moshe Vardi Dec.~17 2022
\end{q}

\Qn Does this sound reasonable to you?

\Ar Totally.
Ron's thesis \cite{Fagin1973} was truly pioneering; the results are published in five papers: \cite{Fagin1974, Fagin1975a, Fagin1975b, Fagin1975c, Fagin1976}.
I have already mentioned his elegant proof of the 0-1 law \cite{Fagin1976}.
Another major result is that connectivity of undirected graphs is not expressible in existential monadic second-order logic \cite{Fagin1975a};
since disconnectedness is expressible in the logic, the collection of expressible properties is not closed under complementation.
The result known as Fagin's Theorem asserts that the expressive power of existential second-order logic on finite structures is exactly that of the NP complexity class \cite{Fagin1974}.
I mentioned above that B\"uchi-Elgot-Trakhtenbrot Theorem was the first descriptive complexity result.
But Fagin's Theorem started descriptive complexity theory in earnest. It wasn't about finite automata but about one of the most important complexity classes, NP.

As far as relational database theory is concerned, the timing of Ron's thesis was perfect.
Codd's 1970 and 1972 papers \cite{Codd1970} and \cite{Codd1972} respectively laid foundations for the relational model of database management, which provided motivation and application for the nascent finite model theory.
In a different letter, of Dec.~17 2022, Ron wrote to me:
``[W]hen I transferred from IBM Watson to IBM San Jose in 1975, Ted Codd became my mentor, which led me to write papers explicitly on relational data base theory.''

In the 1970s and early 1980s there was much DBT work of relevance to FMT.
I know several experts in DBT who are also experts in FMT --- Serge Abiteboul, Phokion Kolaitis, Victor Vianu, and Moshe Vardi; they can tell you all about that.
The book \cite{AHV} is a good DBT textbook for logically inclined.
One DBT article of influence on FMT that will play a role later in our story is the article \cite{CH} by Ashok Chandra and David Harel.

In 1980 Neil Immerman defended his PhD thesis \cite{Immerman1980}; the results were published in \cite{Immerman1981, Immerman1982a}.
I asked Neil what led him to FMT.
\begin{q}
``I was inspired by Fagin's characterization of NP and his proof that connectivity is not in monadic SOE [existential monadic second-order logic],'' replied Neil, ``I felt that lower bounds on full arity SO was going to be too hard, but I thought that first-order lower bounds would be more tractable.  This was what inspired me to work in my thesis area developing what I called "First-Order Expressibility".  Later Hartmanis [Neil's advisor] suggested, "Descriptive Complexity" as a more appealing name.''\hfill --- Neil Immerman Dec.~20 2022
\end{q}

In STOC 1982, Immerman and Vardi presented their versions of what became known as the Immerman-Vardi Theorem; they showed that,  on ordered structures, the least fixed point extension of first order logic captures polynomial time, widely viewed as the most important complexity  class \cite{Immerman1982b, Vardi1982a}.
Immerman went on to become a champion, maybe I should say the champion, of descriptive complexity theory, as both a contributor and an evangelist; see his book \cite{Immerman1999} on the subject.

\Qn What about your contributions that Vardi mentioned?

\Ar I'll speak more about that in the next section.
Here let me mention the very first result:
Under a natural interpretation over finite domains, a function is primitive recursive if and only if it is logspace computable, and it is general recursive if and only if it is polynomial time computable \cite{G051}.

\Qn Did Fagin reply to your second question?

\Ar Yes. He wrote: ``I don't know (and neither does Moshe) who coined the term `finite model theory'.  Moshe says he will look into that.''
Moshe wrote to us the same day and surprised me.
``The first paper that I found that used FMT in its title
is \cite{G090}. Ron followed up with \cite{Fagin1993}.
So it seems that Yuri deserves the credit for coining the term%
\footnote{I took the liberty of replacing URLs with references to the bibliography}.''

\Qn Did you coin the term?

\Ar Not consciously.
Contrary to Moshe and Ron, I did not belong to the database community.
I came from model theory, and the term ``finite model theory'' suggests itself to model theorists working with finite models.
And it is possible of course that the term was used earlier though not in the titles of papers.

\section{The importance of being finite}

\Qn How did you arrive to relational databases?

\Ar At FOCS 1982, Moshe Vardi presented ``On decomposition of relational databases'' where he made essential use of Beth's Definability Theorem
\cite{Vardi1982b}.
I worried that the theorem may not survive the restriction to finite structures and I asked him whether his databases may be infinite. He said yes.

\Qn Infinite databases! This is crazy.

\Ar And a little expensive.

\Qn Right.

\Ar I thought that first-order logic may not be the best fit for databases.

\Qn What gave you this idea?

\Ar My prior experience with ordered abelian groups (OAGs).
In my PhD thesis \cite{G003}, I proved the decidability of the first-order theory of OAGs, solving a problem of Tarski.
But the OAG algebra is rarely first-order. So I extended the decidability to the monadic second-order theory of OAGs where the set variables range over convex subgroups \cite{G025}.
The expanded OAG theory essentially subsumed the OAG literature of the time while the decidability proof became simpler.
Obviously first-order logic wasn't the best fit for OAGs.

Returning home after the FOCS conference, I constructed counterexamples for some famous classical logic theorems, including Beth's Definability, in the finite case, that is the case where only finite structures are taken into account \cite{G060}.
In the process, I discovered that William Tait had given the counterexample for the \L o\'s-Tarski theorem described in \S\ref{sb:SST}.

\Qn Was there a classical theorem that you expected to fail in the finite case but couldn't find a counterexample?

\Ar Yes, Lyndon's theorem \cite{Lyndon}.
It states that a first-order sentence $\phi(P)$ is monotone in $P$ if and only if it is equivalent to a sentence $\psi(P)$ where $P$ has only positive occurrences.
Eventually Miki Ajtai and I proved that it fails in the finite case \cite{G072}.

\Qn Why did Codd strive to achieve the expressivity of first-order logic.

\Ar That was rather natural, I guess.
First-order logic is relatively expressive.
A great many relational queries are expressible in first-order logic.
Besides, first-order logic is the logic of standard logic textbooks.
What else was there?
Propositional logic is way too weak while second-order logic is way too expressive.

\Qn But first-order logic was developed in the framework of foundations of mathematics, I understand, where infinite structures are indispensable.

\Ar This is true.
Also, in the foundational framework, it is important that first-order logic admits an adequate deductive system, where a sentence is provable if and only it is valid.
In fact, first-order logic was introduced as a deductive system. It was obvious that all provable sentences are valid.  G\"odel's completeness theorem established that all valid sentences are provable.
As we saw in \S\ref{sb:trakh}, finite validity is not recursively enumerable. Thus there is no logic calculus where a sentence is provable if and only it is valid on all finite structures.

\Qn In reality, had first-order expressibility proved insufficient?

\Ar Yes. There is no first-order query (or even existential monadic second-order query) that, given an arbitrary relation $R$ returns the transitive closure of $R$ \cite{Fagin1975a}, and transitive closure is needed in database management \cite{Zloof}.

\Qn If first-order expressivity isn't sufficient, what can you do?
Maybe you can carve out an appropriate fragment of second-order logic.

\Ar Yes, but it is more convenient to think in different terms, like transitive closure or least fixed point operators \cite{AU}.

\Qn But what guides you?

\Ar Ah, this is the key point. The answer is computational complexity.

\Qn What's the connection exactly?

\Ar This is a long story. Let me refer you to the paper \cite{G060}.

\section{Metafinite structures}
\label{s:meta}

\Qn You said that you didn't coin the term ``finite model theory'' consciously.
Have you ever consciously coined a term?

\Ar Sure, more than once. One of those terms is relevant to our current conversation: metafinite.

\Qn Hmm, you can't be half pregnant or for that matter half infinite.
You don't claim that databases are metafinite, I presume, as you claimed a minute ago that they are finite.

\Ar I do claim that databases are metafinite.
Let me explain.
A database contains only finitely many records (tuples in relations) at any given moment.
In that sense it is perfectly finite.
It is also dynamic as we saw in \S\ref{s:codd}.
But implicitly databases often involve static infinite structures.

\Qn We considered a salary database in \S\ref{s:codd}.
Does it involve an infinite structure?

\Ar Yes, a query to a salary database may return a number, say the average salary, that does not appear in the database.
Where does this new number come from?
From the infinite structure of rational numbers.

\Qn That involvement of an infinite structure, what is it mathematically?

\Ar A good question. Erich Gr\"adel and I worked on it \cite{G109}.
In our formalization, a metafinite structure consists of
\begin{enumerate}
\item a primary part, which is a finite structure,
\item a secondary part, which is a (usually infinite) structure, e.g.\ the arithmetic of real numbers, and
\item ``weight''\ functions from the primary part to the secondary.
\end{enumerate}

\Qn I guess I see how the salary database looks as a metafinite structure.
The primary part is the set of employees, the secondary part is  rational arithmetic, and the unique weight function assigns employees their salaries.
But how do you compute the average salary?

\Ar In this example, the secondary part would be an extended rational arithmetic, where the arithmetical operations are applied to finite multisets.
In particular, one can obtain the total salary of all employees and the number of employees.

\Qn How is this related to finite model theory?

\Ar Quantification is allowed over the primary part only.

We investigated the theory of metafinite structures, with Erich doing most of the mathematical work.
We found out that many results of finite model theory survive, sometimes in more than one form, in metafinite model theory.

\Qn Do you view metafinite model theory as a separate research area?

\Ar No. In my view, it belongs to the research area known as finite model theory.

\section{Genericity}\label{s:gen}
%AHV IX, 419

\Qn I presume that tailoring logics to complexity, at least to polynomial time complexity, has been achieved.

\Ar Not really. A complication arose.
The traditional model of computation used to define complexity classes is the Turing machine model.
Inputs for Turing machines are strings.
Your algorithm may work with graphs or other structures, but you have to encode your input structure as a string to produce input for a Turing machine.
We need a more general computation model that works directly with structures, and replacing your input structure with an isomorphic one should not affect the computation.

\Qn This looks to me like a generalization that only mathematicians can come with.
What on earth has it to do with query languages?

\Ar In database theory parlance, query languages should be \emph{generic} which means that they should abstract from the physical layer of databases.
In this connection, let us call more general computation models that work directly with structures \emph{generic}; also let us call polynomial time defined according to a generic computation model \emph{generic}.

\Qn What does it mean to abstract from the physical layer?

\Ar For simplicity, consider a database that holds just one relation.
That relation is a set of tuples.
There is no order on the set.
Accordingly, a query asking for the first tuple is meaningless.
On the other hand, in a memory of a computer hosting the database the tuples are stored in some order.

\Qn Fine, take advantage of that order and increase the expressivity of your query language.
Specifically, the first-tuple query will return a particular tuple.

\Ar But the order is accidental. It varies from one database host to another.
The first-tuple query may return different tuples on different hosts.

In \cite{CH}, Chandra and Harel raised an important question.
Fix some standard encoding of structures by binary strings and call a Turing machine $M$ \emph{generic} if the set of structures $S$ such that $M$ accepts the standard string encoding of $S$ is closed under isomorphisms.
Chandra and Harel asked whether the set of generic Turing machines is recursive.
If the answer is positive, then the good Turing machines constitute a generic computation model.

In \cite{G074}, I conjectured that the answer to the Chandra-Harel question is negative and that no reasonable (satisfying some minimal conditions) logic captures generic polynomial time on structures.
The question remains open.

Actually, the situation is a bit strange.
In \cite{G120,G150} a logic is defined --- it is called BGS (after the authors) --- such that every BGS definable property of structures is generic polynomial time computable and we don't know any generic polynomial time property that is not BGS definable.

\Qn Do you believe that BGS captures generic polynomial time?

\Ar No, most probably it does not.
In \cite{Rossman}, Ben Rossman came close to exhibiting a set of structures that is computable in generic polynomial time but is not BGS definable. He constructed a generic polynomial time computable function that (a)~given a finite vector space, constructs its dual and (b)~is not BGS definable.

\section{Postlude}\label{s:post}

\Qn Have you now covered all DBT/FMT developments that you witnessed or participated in?

\Ar No. One issue is logic programming languages, specifically Prolog and Datalog.
That issue is involved and richly deserves a separate conversation.

\Qn Did finite model theory cut the umbilical cord to database management?

\Ar Yes. ``DBT was a huge motivator for FMT,'' wrote Moshe Vardi to me on Nov.~24 2022, ``but then FMT moved from PODS [the Symposium on Principles of Database Systems, the premier international conference on the theoretical aspects of database systems] to LICS [ACM/IEEE Symposium on Logic in Computer Science], which is a conference about mathematical foundations.''

\Qn Is finite model theory still a separate research area?

\Ar Research areas are social groups to an extent.
My impression is that finite model theory becomes less of a separate research area in spite of strong internal social/communal connections.
But I am not the right person to answer this question as I had moved on to other research areas quite a while ago.

\subsection*{Acknowledgement}
As I worked on this little paper, I exchanged letters with Anuj Dawar, Ron Fagin, Erich Gr\"adel, Neil Immerman, Mark Liogonky, Ben Rossman, and Moshe Vardi.
Their timely replies were most helpful.
And I much benefited from discussing all aspects of this paper with Andreas Blass.
%My wholehearted thanks to Andreas, Anuj, Ron, Ben, and Moshe!
Finally, the photo of Boris Trakhtenbrot with his students comes courtesy of the Trakhtenbrot family archives, the photos of Dana Scott and William Tait come courtesy of Scott and Tait respectively, the photo of Patrick Suppes is from the Guggenheim Foundation, %https://www.gf.org/fellows/all-fellows/patrick-suppes/
and the photos of Yuri Glebsky and Edgar Codd are from Wikipedia.

\end{document}